\documentclass[journal,10pt]{IEEEtran}

\usepackage{graphicx}
\usepackage{caption}
\usepackage{subcaption}
\usepackage{tabularx}

\emergencystretch=\maxdimen
\begin{document}

\title{Analyzing Contact Patterns in Public Transportation Systems for Opportunistic Communication Services}

\author{Eduardo R. Manika, Emilio C. G. Wille, Joilson Alves Jr
\thanks{E. R. Manika, E. C. G. Wille, J. Alves Jr, Federal University of Technology -- Paran\'{a} (UTFPR), Av. Sete de Setembro 3165, 80230-901, Curitiba (PR),  Brazil. E-mail: manika@utfpr.edu.br, ewille@utfpr.edu.br, joilson@utfpr.edu.br}
}

\maketitle

\begin{abstract}
Vehicle mobility has a significant impact on wireless communication between vehicles (buses) in \emph{Public Transportation Systems} (PTS).
Nevertheless, the transportation literature does not provide satisfactory models for bus movements because they are influenced by a variety of factors (itineraries, timetables, etc.). Custom-made mobility models that take these issues into account require a great deal of effort and may render simulations unfeasible. 
This article considers  a tool (EMMS) that automatically inserts PTS information into a mobility simulator in order to undertake a complete statistical analysis of vehicular density, trip duration, and vehicle-to-vehicle interaction. In light of opportunistic communication services, this analysis is of the utmost importance.
\end{abstract}

\begin{IEEEkeywords}
 Smart cities, Public transportation system, Geopositioning data, Map-matching, Statistical analysis, VANETs, SUMO.
\end{IEEEkeywords}

\section{Introduction}

\emph{Smart Cities} have been standing out in recent years through the use of \emph{Information and Communications Technology} (ICT). One of the main goals of smart cities is a higher efficiency in public administration, in addition to minimizing urban problems, thus creating a more sustainable and better city to live in \cite{FOUNOUN2018,ZANDBERGEN2017}. The setting up of smart cities is strongly related to the collection and prior analysis of urban data, aiming at assessing how cities behave and develop. In possession of these data, various technologies and applications are being developed to help the population with their daily routines, to improve the mobility of people and vehicles, as well as to reduce negative environmental impacts, among others, \cite{Appio19,Alawadhi2012,Ijemaru22}. 
The most critical components of smart cities are public transportation and urban mobility \cite{Kozievitch2016}. Intelligent solutions making use of wireless communication can contribute to the solution of these problems, improving transportation mobility and safety in large centers, in addition to providing applications that enable traffic control and management. These solutions and applications make up the \emph{Intelligent Transportation Systems }(ITS) \cite{Garg2022}.

In a \emph{Public Transportation System} (PTS), vehicles (buses) travel between stops according to a predetermined schedule. Each vehicle has a itinerary, which is the ordered sequence of stops that the vehicle traverses. A bus line corresponds to a set of different vehicles with the same itinerary.
Understanding mobility of vehicles in PTSs is of utmost importance. For example, the possibility of wireless communication between vehicles is strongly influenced by the vehicle (node) mobility \cite{JoilsonETT16,wille16}. However, the transportation literature does not provide acceptable models for bus movements in an urban environment since they are affected by vehicular and passenger traffic conditions, lanes organization, traffic signal management, company policies, and others. 

In the context of communication between vehicles, there are the \emph{Vehicular Ad Hoc Networks} (VANETs), a type of network that has been exerting great influence on the ITS scenario \cite{Maimaris16,Joilson2015,Coutinho15,Zhang13}. 
Design and development of VANETs require feasibility, implementation and operation studies. Among those studies it should be emphasized the ones based on simulation, due to their low operational and financial costs compared to the real in-field mobility studies. 
Among the mobility simulators frequently mentioned in literature, it is noteworthy the \emph{Simulator of Urban MObility} (SUMO) \cite{SUMO}.
An important feature of SUMO is the possibility of manual creation of simulation maps and the import of real maps (road scenarios with data previously obtained from other services, such as \emph{OpenStreetMap}). SUMO easily integrates with various network simulators, such as NS-2 and Omnet++.

It should be noted that buses belonging to the public transportation system can be seen as nodes of a kind of opportunistic data network. The model that represents the behavior of PTS vehicles can be inserted into the SUMO. However, this model has its particularities, such as the following: route characterization (itineraries); departure timetables; arrival at specific points of the itinerary, and bus stops along the way. It is possible to manually generate this model (itineraries, timescales, etc.) to serve as input into SUMO. However, this process is very costly and can make simulations impractical.

This article conducts a detailed statistical analysis of data from geolocation records of public transportation buses in the city of Curitiba (a city in southern Brazil, well-known by its bus-based transportation system) obtained through the SUMO simulator. With that on mind, a computational tool was developed in order to automate the maps configuration, importing the actual characteristics of public transportation to SUMO. This tool, named \emph{Engine for Map-Matching to SUMO} (EMMS), imports in an automated manner the following characteristics of the public transportation system to the simulated environment: bus lines routes, locations of bus stops and departure times. That allows for more agile, reliable and realistic simulations \cite{mwa24}. The results obtained from this investigation bring about information related to vehicular density, travel time, and statistics of contact between vehicles (considering two regions of Curitiba), which are of great importance within the ITS context. 

\IEEEpubidadjcol 
The remainder of this paper is structured as follows. 
Section \ref{sec:2} gives a review of related work.
The description and implementation of EMMS are given in Section \ref{sec:3}.
Section \ref{sec:4} presents the urban scenarios, the validation of EMMS, and the metrics considered in this study.
Section \ref{sec:5} presents and discusses the numerical results obtained through computational simulations.
Finally, Section \ref{sec:6} summarizes the main conclusions.

\section{Related Work}
\label{sec:2}

One of the challenges related to vehicular network studies is the definition of a mobility model that can provide an accurate and realistic description of vehicular mobility. Mobility models are classified as \emph{macroscopic} or \emph{microscopic}, according to the level of details they provide. The macroscopic model considers a more general approach, dealing with characteristics related to flow, such as lane speed and density. On the other hand, the microscopic model focuses on characteristics about vehicle behavior, such as its acceleration and deceleration \cite{Harri2009}.

One of the first mobility models used in vehicular networks is the \emph{Random Node Movement}. This model is characterized by its randomness, both of speed and locomotion, and it does not faithfully replicate the VANETs characteristics \cite{Tian20}. 
Some others examples of randomness-based models are the following: \emph{Random Walk}, \emph{Random Way-point} and \emph{Random Direction}  mobility models \cite{Kumar12}. In order to realistically represent the VANETs characteristics, another model was developed, called \emph{Artificial Mobility Traces} \cite{CAMP2002}, which in addition to reproducing the simulation environment of VANETs, also offers parametrization freedom (varying the rate of acceleration, speed and randomness). This model was divided into two classes, depending on the level of detail in the model creation, which are: \emph{Traffic Stream Models}, characterized as macroscopic, as it bases vehicular traffic on three variables: speed, density, and flow; for that reason they are less common in simulations. The other class is the \emph{Car-following Model}, characterized as microscopic, and it is based on four variables: acceleration rate, deceleration rate, maximum speed and degree of randomness.

In general, it is necessary that the mobility model used in the simulations reproduce in the most realistic way the area to be analyzed, taking into account the local characteristics. A realistic mobility model should consider the following characteristics \cite{Ahmed2006}: \\

\begin{itemize}
\item Realistic topological maps;
\item Existence of obstacles;
\item Departure and arrival points;
\item Acceleration and braking;
\item Day periods;
\item Non-random vehicle distribution;
\item Driving behavior. \\
\end{itemize}

 The study conducted in \cite{Gaito2010} investigates whether a public transportation network used as a VANET backbone can be a viable solution, as it presents a large coverage area and pre-programmed paths (itineraries) repeatedly run. The goal of this work was to compare the routing performance of a distance-vector when applied to a more realistic environment. Experiments were carried out using data of the actual topology of the city of Milan (Italy). A large number of nodes were simulated at actual timetables of the public transportation.

Another issue concerns the import of real data to the simulation environment. In this context comes the \emph{map-matching} method, which consists of identifying a match in a sequence of points, for example, coordinates of geolocation to a digital map \cite{Quddus2007}.
The study conducted in \cite{Quddus2007} shows a literature review of real-time and post-processing map-matching methods, categorizing the approach of algorithms for this purpose in four groups: geometrical, topological, probabilistic and advanced techniques. The study summarizes the main limitations and constraints of the existing algorithms, from the simplest search algorithms (point-to-point correspondence, point-to-curve correspondence) to more complex approaches, including probability theory applications, fuzzy logic and belief theory.
The authors in \cite{Raymond16} study the problem of identifying vehicle trajectories from the sequences of noisy geospatial-temporal datasets.  They propose a simple and robust technique based on the combination of map-matching, bag-of-roads, and dimensionality reduction for their route identification. Experiments on datasets of buses in the city of Rio de Janeiro (Brazil) confirm the high accuracy of the method.
The methodology proposed in \cite{Perrine15} uses data from \emph{Generalized Transit Feed Specification} (GTFS) files to create a graph representation compatible with an underlying roadway network  graph. The  algorithm  can  assume  that  GPS  points  are  fed  one  at  a  time  and maintains candidate paths between suggested GPS points. This procedure, in combination with the use of an error region around matched points, enables the algorithm to successfully handle GPS and network data of varying quality. Experimental results suggest that the approach is highly successful even when the underlying roadway network is not complete.

Due to the particular characteristics of the movement of the buses belonging to a PTS, the randomness-based models described above turn out inappropriate. The solution consists of simulating bus lines using SUMO, which enables for the simulations from the simplest networks to more complex environments. The micro-mobility presented in SUMO offers a collision-free system, in which a vehicle speed is determined by the speed of the vehicle straight ahead. In a scenario with pathways with more than one lane, this tool accepts the action of overtaking. 

\section{The EMMS approach}
\label{sec:3}

In order to conduct a statistical analysis of data from the Curitiba's bus lines, we resort to the EMMS tool.
As stated, the \emph{Engine for Map-Matching to SUMO} (EMMS), is a software tool that 
 automatically imports the characteristics of the real vehicle mobility environment to the simulated environment (SUMO), including the routes of bus lines, the location of bus stops and the departure times. The EMMS was developed using the Python programming language \cite{Python2018}, which allows the integration with SUMO through a set of programming libraries called \textit{SUMOLib} \cite{SUMO} and is divided into four architectural layers: 

\begin{itemize}
    \item First Layer: Data entry and processing;
    \item Second Layer: Matching of bus itinerary maps;
    \item Third Layer: Matching of bus stops maps;
    \item Fourth Layer: Configuration files required for SUMO. 
\end{itemize} 

\subsection{First Layer: Data entry and processing}
\label{subSec:EntradaDados}

In this layer, data are entered and processed. The mandatory data are: \\

\begin{itemize}
    \item Unique bus line identifier (identifying code);
    \item File containing the topology of the area considered for simulation (road network map), in the format already converted for SUMO (\textit{net.xml});
    \item File containing the route coordinates of the bus in GPS format, with latitude and longitude information;
    \item File containing the bus line departure timetables; 
    \item File with the location of the bus line stops. \\
\end{itemize}

The input files must be in JSON format (text format), following a predefined pattern. After the validation of the mandatory parameters, the conversion of itinerary positioning and bus stops data is carried out, from  geographical coordinates (GPS - latitude and longitude) to  Cartesian coordinates ($X$ and $Y$). In this conversion, SUMO libraries are used, taking into account the file loaded into the data entry, containing the area topology. In order to increase the accuracy and improve efficiency in matching maps, EMMS doubles the amount of positioning points for bus routes provided in the data entry. This duplication increases the accuracy of matching, since urban regions have a high density, and in some situations, positioning points (GPS) provided present a great dispersion. Such duplication is carried out through the calculation of the midpoint between two Cartesian coordinates. As a result, this layer generates a conference file representing bus line routes (with the data already duplicated). This file will be used by the other EMMS layers. 

\subsection{Second Layer: Bus itinerary matching}
\label{subSec:MapMatchingItinerario}

The second layer deals specifically with matching bus itinerary (routes). This itinerary is the trajectory of vehicles (buses) on a particular bus line. In order to get this trajectory in a virtual environment (such as SUMO) it is necessary to detect the route segments (\emph{edges}) in a digital environment, which correspond to GPS coordinates (latitude and longitude) for that route.

\begin{figure}[h]
\centering
\includegraphics[width=0.45\textwidth]{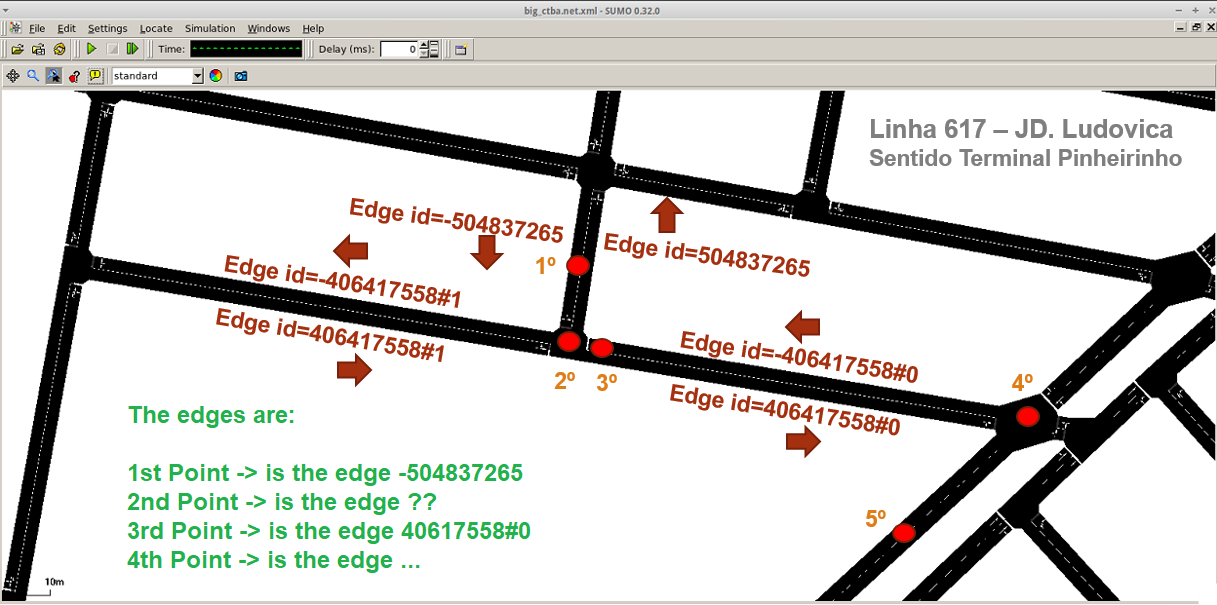}
\caption{Map-matching example (route edge determination).}
\label{fg_simulador01}
\end{figure}

Figure \ref{fg_simulador01} graphically exemplifies the matching of SUMO maps for a particular bus line. The red dots are positioning data (GPS) of the buses, representing the itinerary of a particular bus line. Each of the \emph{edges} on the map has a unique identifier (\emph{id}), which represents a street segment and can be formed by one or more pathways (called \emph{lanes}). The map-matching consists of identifying which \emph{id} corresponds to the positioning data point. Finally, a list of chained edges is generated (represented in green in the figure), which make up the bus trajectory, i.e., the bus line itinerary within the simulation environment that corresponds to the actual data. This chain of edges depends on a ``physical" (interconnection between edges) and ``logical" (related to the direction of edges) interconnection.

The physical interconnection refers to the connection of the edges in the digital map, i.e., the end of a selected edge must be the beginning of the subsequent edge. In SUMO this interconnection is validated through identifiers called \emph{nodes}, which represent the ends of an edge. The logical interconnection, on the other hand, refers to the direction and conversion permissions of the edge, i.e., it represents traffic signs. In SUMO the nodes are used to identify the direction of the edges. During the edge configuration each node receives a unique identifier, the source node (called \emph{from}) and the target node (called \emph{to}) are specified. The configuration of conversion permissions are performed through the element called \emph{connection}, which references the allowed edges.

In order to automate the matching of the bus itinerary maps, EMMS takes the following steps:  \\

\noindent
\emph{Listing of all candidate edges (Step 1):}
For each entry positioning point (GPS), the algorithm determines a region of interest. Such region corresponds to a perimeter with a known radius (predefined in EMMS in 15 meters). The edges within this region are selected as candidates \cite{Hashemi2014}. The Euclidean distance between the positioning point and the line drawn between the nodes (``from" and ``to") of the edges is used to accomplish this discovery. \\

\begin{figure}[!tbh]
\centering
\begin{subfigure}[t]{.28\textwidth}
	\centering
\includegraphics[width=\textwidth]{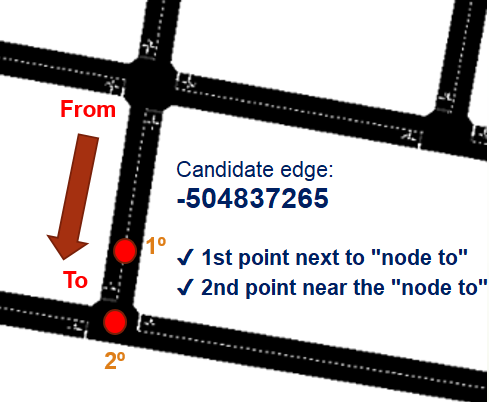}
\caption{First case.}
\label{fg_matching_inicial_01}
\end{subfigure}
\qquad
\begin{subfigure}[t]{.28\textwidth}
	\centering
\includegraphics[width=\textwidth]{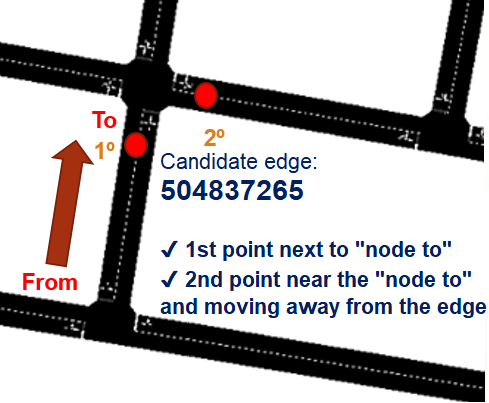}
\caption{Second case.}
\label{fg_matching_inicial_04}
\end{subfigure}
\caption{Analysis for matching the first edge.}
\label{fig:fg_matching_inicial}
\end{figure}

\noindent
\emph{Matching of the first edge (Step 2):}
The matching of the first edge is important for the whole map-matching process. In this situation there is no previous reference point (like another edge), thus requiring a specific treatment. Most of the times it is not possible to separately identify the initial direction the bus will move, so it is necessary to consider the next positioning point. It is known that the bus moves from the first towards the second positioning point and that the sequence is always from the node ``from" (origin) to the node ``to" (destination). Therefore, it is possible to correctly map the first edge, by validating whether the edge direction corresponds to the actual direction of the vehicle. One way of analyzing the edge direction is by checking the Euclidean distances between positioning points (first and second) against the nodes (``from" and ``to") of the candidate edge selected. In general, EMMS carries out the weighing of the initial matching, first selecting the shortest distance edge among the candidate ones. Subsequently, EMMS verifies whether the  positioning points are approaching or moving away from the candidate edge node. If the selected edge does not satisfy any of the above mentioned conditions, the selected edge is discarded and the process restarts with the next candidate edge. Figure \ref{fg_matching_inicial_01} shows one of the criteria analyzed, and if the second positioning point is closer than the first one to the node ``to", this represents a high probability that the edge corresponds to the real environment, because the node ``to" is always the final destination of the edge.

Another situation of correct matching is presented in Figure \ref{fg_matching_inicial_04}, in the case that the first positioning point is closer to the node ``to" and the second positioning point is moving away from the node ``to", also on another edge. As a result of this step, EMMS returns the first matched edge, which will be used to define the next matches.

\noindent
\emph{Matching of the next edge, considering the previous one (Step 3):}
This step consists of matching all the other positioning points, considering the edge selected in the matching process of the immediately preceding positioning point. It is noteworthy that the dependence between the edge connections, both physical and logical, is one of the operational characteristics of SUMO. Both conditions need to be met so that the simulation will be performed successfully. For this reason the manual process of matching is costly, sometimes making the generation of simulations unfeasible. EMMS, on the other hand, enables the automation of these matches, performing the necessary validations (both physical and logical), making the whole process more agile.

Thus, the logic of this stage is divided into two phases: firstly, the candidate edge with the shortest distance to the positioning point is retrieved and subsequently the physical and logical interconnections between the edges (candidate and selected in the previous process) are  analyzed. The validation of the physical interconnection is carried out by comparing the \emph{ids} of node ``to" of the matched edge of the previous positioning point and the node ``from" of the selected candidate edge. Once this step is validated, the analysis of the logical interconnection is performed checking for connection registration on the digital map (\textit{net.xml}). If it does not exist, it means that although there is a physical connection between the edges, the continuity towards the candidate edge is not permitted, or conversion is not allowed, therefore the selected candidate edge is discarded and the process restarts, seeking the next candidate edge. If the logical interconnection exists (has a record), the edge is added to the chained list of edges and it will be used in the matching of the next positioning point.
At the end of the matching process, the sequence of edges that represents the itineraries of the buses in the real environment will be ready to SUMO. In addition, a file for manual conference will be generated.

\subsection{Third Layer: Matching the bus stop points}
\label{subSec:MapMatchingPontosOnibus}

This layer consists of matching the bus stop points to the corresponding edges of the itinerary matching process carried out in the second layer, as well as locating the bus stop along the edge matched.
Firstly, the EMMS loads the bus stop location data (provided in the first layer). Subsequently, it carries out a search of the candidate edges within the region of interest, close to each bus stop point indicated, considering the same radius configured at the data entry.
After the candidate edges search, an analysis of the correspondence between the candidate edges and the edges mapped in the bus line route matching process (second layer) is performed.    
If there is a correspondence between them, the bus stop is matched. The configuration of the bus stop location within the edge is performed by calculating the (Euclidean) distance from the bus stop to the node ``from" of the corresponding edge. Thus, it is possible to identify the starting point of the bus stop in the virtual environment.

\subsection{Fourth Layer: Creation of configuration files for running a simulation using SUMO}
\label{subSec:CriacaoArqConfSUMO}

The fourth layer is responsible for automating the generation of configuration files required to run simulations in SUMO. In that process, are generated two files: one responsible for setting up the buses in the simulation (\textit{rou.xml}), and the other responsible for configuring the bus stops for the simulation environment, in addition to configuration of the specific location of the stop point inside the edge (\textit{add.xml}).

\section{Analysis of Vehicle Mobility in a Public Transportation System}
\label{sec:4}

This section presents the urban scenarios considered, the validation of the mobility simulation environment, and the metrics considered in this study.
The simulations were carried out using the actual data of geolocation of the public transportation buses of the city of Curitiba (these data were made available by the public transportation company – URBS \cite{WSURBS2018}).
The data analyzed refer to October 25th, 2022, from 4:00 pm to 8:00 pm. The data import to SUMO was performed through EMMS. Initially, data from two city regions were extracted. The first corresponds to the central region, where 21 bus lines depart from \emph{Rui Barbosa Square}, totalizing 100 buses. 
The second corresponds to the southern region, where 18 bus lines depart from \emph{Pinheirinho Terminal}, totalizing 52 buses. 

For the validation of the mobility simulation environment, compatibility analysis were carried out considering the actual journey times of the buses (made available by the URBS) and simulated travel times (considering the trace mobility generated by SUMO). In general, the central area (Rui Barbosa) presented a compatibility percentage of 95.2\%, while in the southern region (Pinheirinho Terminal) it was 89.2\%, during the considered period (afternoon). An evaluation considering morning hours revealed that the compatibility in the central region was 90.5\% and in the southern region it was 91.6\%. Thus, that high degree of compatibility in varied cases ensures that the results obtained from the simulation environment are reliable.

\subsection{Mobility Metrics}
\label{sec:metricasSimulacao}

The mobility metrics (described below) regard physical issues, i.e., they measure the spatio-temporal relation between vehicles or between vehicles and the environment. These metrics are key to studying the feasibility of establishing a data network, considering the buses belonging to the urban transportation network as active elements.

The first two metrics enable the measurement of the vehicles displacement time and the vehicle density. The three following metrics aim to measure the connectivity between vehicles. It is considered that two vehicles will be in contact (or connected) when the physical distance between them (inter-vehicular distance) is shorter than or equal to a previously established value (in this case, the transmission radius of the wireless communication system is considered). That way, one can measure the contact and inter-contact times between vehicles, as well as the number of vehicles that have contact inside the transportation network. \\

\noindent
\emph{Travel Time:} 
It represents the time a vehicle (bus) takes to arrive to a reference point (bus terminal), from the moment its distance to the reference point is less than a pre-established value.  \\

\noindent
\emph{Total Number of Vehicles in the Perimeter:} 
It is the number of vehicles that are within a pre-established perimeter, around a reference point (bus terminal) at any given time. This metric allows one to roughly calculate the vehicular density. \\

\noindent
\emph{Total Number of Connected Vehicles:}
It corresponds to the total amount of vehicles connected at any given time considering the whole region of interest.  \\
 
\noindent
\emph{Vehicle Connectivity:} 
The vehicle connectivity of a given vehicle corresponds to the number of vehicles it keeps a contact with at any given time. That metrics aims to assess the amount of buses with which a vehicle can communicate, thus creating possible transmission networks. \\

\noindent
\emph{Contact Time and Inter-contact Time:} 
The contact time corresponds to the time interval during which a given vehicle  remains in contact with another one or more vehicles. The inter-contact time corresponds to the time elapsed between an interruption and the reestablishment of a new connection with any other vehicle. 
The contact time is a relevant factor affecting the amount of data that can be transferred
between nodes when they come into range. Meanwhile, the inter-contact time affects the frequency with which 
packets can be transferred between networked devices. 
 
\begin{figure}[h]
\centering
\includegraphics[width=0.45\textwidth]{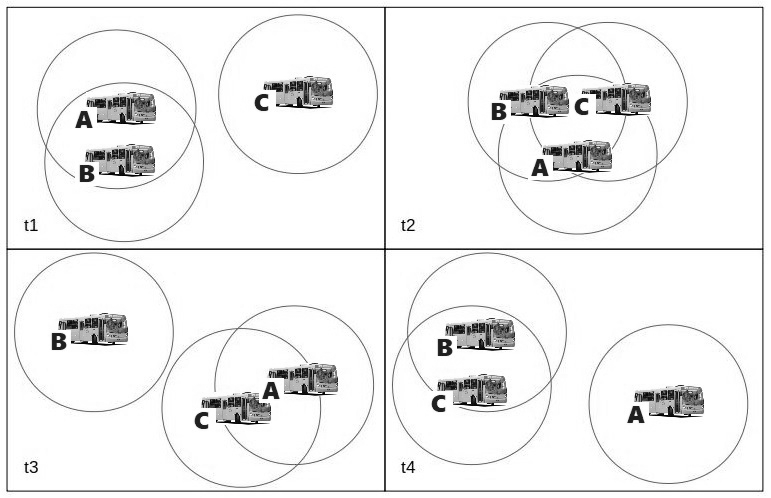}
\caption{Calculation of Contact Time and Inter-contact Time example.}
\label{fig:figuraExemploTempoConexao}
\end{figure}

Figure \ref{fig:figuraExemploTempoConexao} shows an example containing four consecutive sampling moments (\textit{t1}, \textit{t2}, \textit{t3} and \textit{t4}) and the buses (\textit{A}, \textit{B} and \textit{C}) positioning in each moment. In this example, vehicle A would have contact time equal to $t4 - t1$, being \textit{t4} the beginning of the period without connection. Bus \textit{B} would have contact time equal to $t3 - t1$ and inter-contact time equal to $t4 - t3$ being \textit{t4} the beginning of the connected interval.

\section{Simulation Results and Discussions}
\label{sec:5}

In this section, the results of simulations to assess the behavior of vehicular mobility are presented. The analysis are performed through graphs known as \emph{violin-plot}  \cite{Hintze1998}, which correspond to a combination of \textit{boxplot} (box diagram) and \textit{density trace} (smoothed histogram).
The boxplot is a graph that simultaneously gathers important information from the data set such as the dispersion or variability, the maximum and minimum values, the median and discrepant observations in the data (called \emph{outliers}). The boxplot divides the data set into four parts, each representing approximately 25\% of the total, called \emph{quartiles}.
The density trace is presented symmetrically to the left and right of the boxplot, allowing one to evaluate the distribution of the data set, highlighting the peaks, valleys and bumps in the distribution.

\begin{figure}[h]
\centering
\begin{subfigure}[t]{.35\textwidth}
	\centering
	\includegraphics[width=\textwidth]{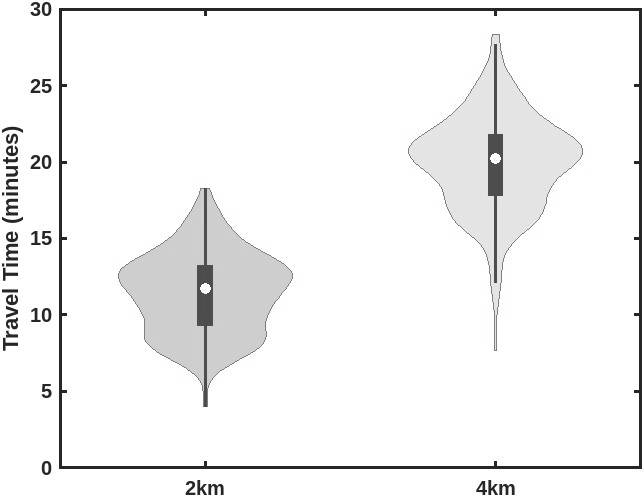}
	\caption{Rui Barbosa Square}
	\label{fig:subfigureTempoChegTerm_a}
\end{subfigure}
\qquad
\begin{subfigure}[t]{.35\textwidth}
	\centering
	\includegraphics[width=\textwidth]{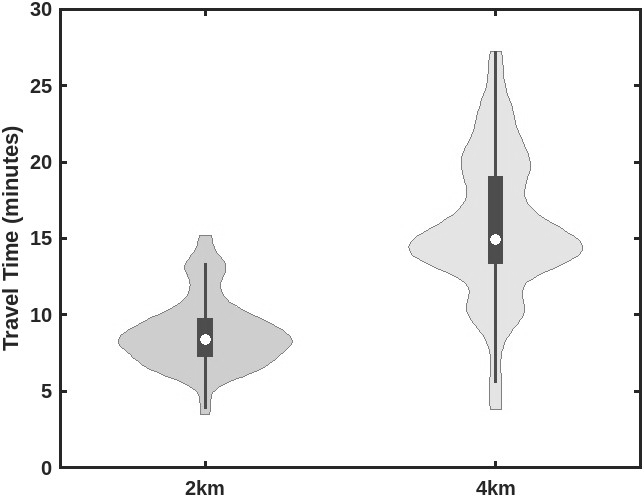}  
	\caption{Pinheirinho Terminal}
	\label{fig:subfigureTempoChegTerm_b}
\end{subfigure}
 \caption{Travel Time for two regions of Curitiba (afternoon).}
\label{fig:subfigureTempoChegTerm}
\end{figure}

In order to evaluate the metrics, data sampling was performed every 3 seconds. For the determination of contacts, transmission range of 150 m and 300 m were considered \cite{AlMEIDA2018}. The reference points considered were Rui Barbosa and Pinheirinho terminals, 
and the perimeter (under analysis) around reference points has a radius of 2 km or 4 km. \\

\noindent
\emph{Travel Time:}
Figure \ref{fig:subfigureTempoChegTerm} shows the vehicle travel time towards the terminals considering the two regions of the city (central and south) in the afternoon period. Perimeters with radius of 2 km and 4 km were considered.  
The travel times have similar values, with a slight superiority of the central region (Rui Barbosa - top plot) 
compared to the southern region (Pinheirinho - bottom plot), 
especially with a 4 km range. This is because the central region has a larger number of traffic lights around the terminal (the traffic lights are also imported into the simulation map) and also due to more buses traveling in this part of the city, generating more traffic congestion. It can be observed that in the central region and with a 2 km radius, the travel time (mean value)
is close to 12 minutes and to the southern region, it is close to 9 minutes. With a 4 km range, the travel time in the central region is around 20 minutes, while in the southern region it is about 16 minutes. 

We note that, from these distributions, the behavior of the vehicle speed can be easily determined and, in general, it
follows approximately a normal distribution. \\

\begin{figure}[h]
\centering
\begin{subfigure}[t]{.35\textwidth}
	\centering
	\includegraphics[width=\textwidth]{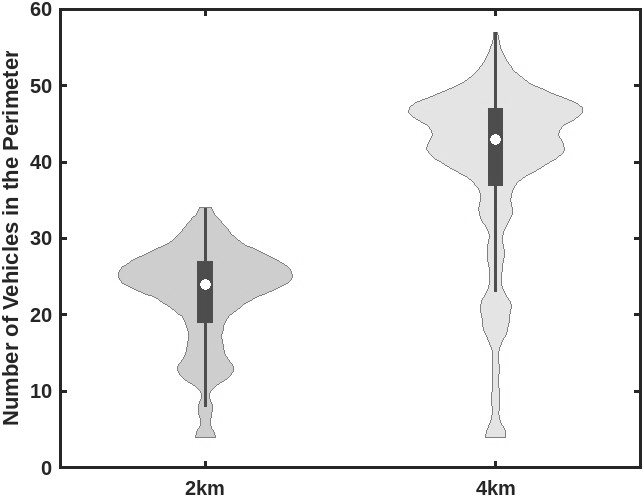}
	\caption{Rui Barbosa Square}
	\label{fig:subfigureDensidadeTerm_a}
\end{subfigure}
\qquad
\begin{subfigure}[t]{.35\textwidth}
	\centering
	\includegraphics[width=\textwidth]{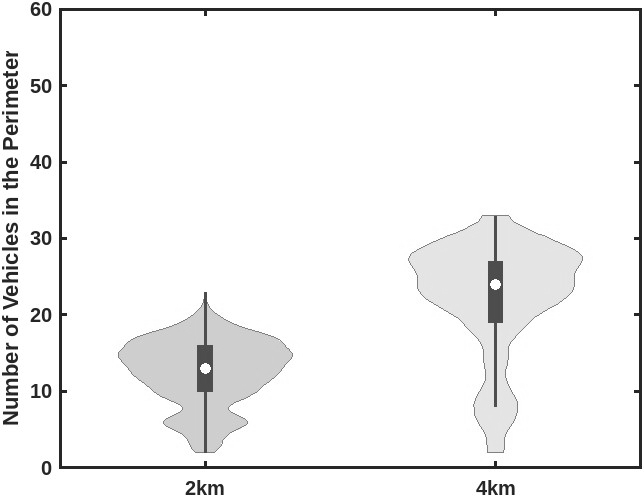}  
	\caption{Pinheirinho Terminal}
	\label{fig:subfigureDensidadeTerm_b}
\end{subfigure}
\caption{Total Number of Vehicles in the Perimeter.}
\label{fig:subfigureDensidadeTerm}
\end{figure}

\noindent
\emph{Total Number of Vehicles in the Perimeter:}
Figure \ref{fig:subfigureDensidadeTerm}  presents the results obtained for the distribution of buses near the terminals (Rui Barbosa and Pinheirinho) in the afternoon. Perimeters with radius of 2 km and 4 km were considered.  It is observed that Rui Barbosa Square (top plot)
has a higher number of buses within the area considered, compared to Pinheirinho Terminal (bottom plot)
as expected, due to the greater number of bus lines present in the central region and, consequently, a larger number of buses. The violin-plot format indicates a larger amount of buses near the upper limit. The  \emph{interquartile range} is relatively small in all cases, indicating a relative stability of the bus lines during the analyzed time interval. \\

\begin{figure}[h]
\centering
\begin{subfigure}[t]{.35\textwidth}
	\centering
	\includegraphics[width=\textwidth]{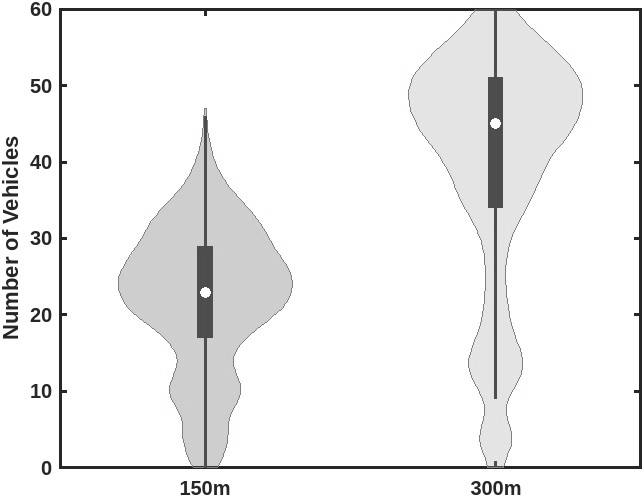}
	\caption{Rui Barbosa Square}
	\label{fig:subfigureFatorCont_a}
\end{subfigure}
\qquad
\begin{subfigure}[t]{.35\textwidth}
	\centering
	\includegraphics[width=\textwidth]{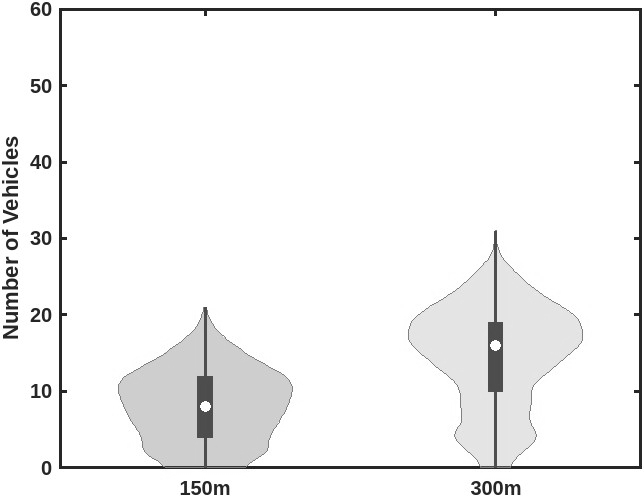} 
	\caption{Pinheirinho Terminal}
	\label{fig:subfigureFatorCont_b}
\end{subfigure}
\caption{Total Number of Connected Vehicles for two regions of Curitiba (afternoon).}
\label{fig:subfigureFatorCont}
\end{figure}

\noindent
\emph{Total Number of Connected Vehicles:}
Figure \ref{fig:subfigureFatorCont} shows the total number of connected vehicles, i.e., the total number of buses connected across the simulation scenario, with at least one bus inside the transmission range. The two regions of the city were analyzed, considering the  transmission range of 150 m  and 300 m. It can be noticed that in the city central region (top plot)
the total amount of buses connected throughout the simulation was higher than the data presented in the southern region (bottom plot)
as expected, due to the lower number of buses in circulation in this region (100 buses in the central region and 52 in the southern region). It should be mentioned that in the central region there was a maximum of 46 buses connected (not necessarily all of them to each other, but all with at least one neighbor) (150 m transmission range) and 72 buses (300 m transmission range). \\

\begin{figure}[h]
\centering
\begin{subfigure}[t]{.35\textwidth}
	\centering
	\includegraphics[width=\textwidth]{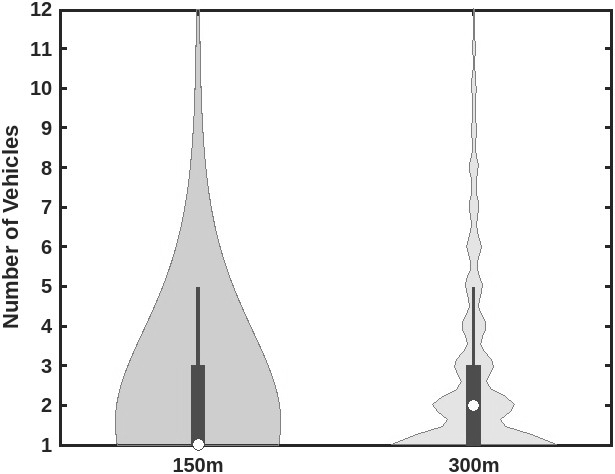}
	\caption{Rui Barbosa Square}
	\label{fig:subfigureQtdeCont_a}
\end{subfigure}
\qquad
\begin{subfigure}[t]{.35\textwidth}
	\centering
	\includegraphics[width=\textwidth]{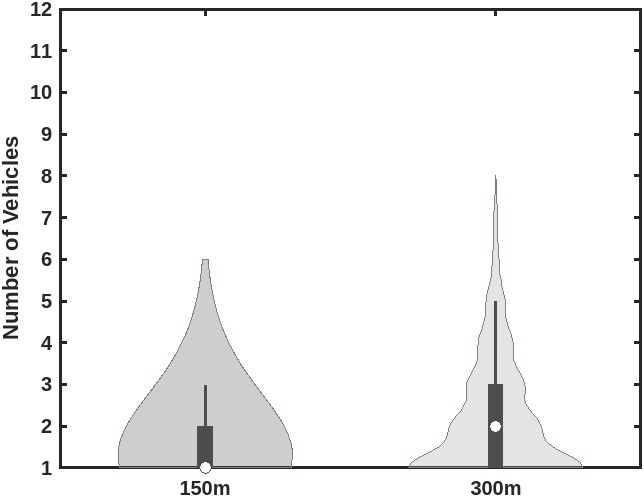}  
	\caption{Pinheirinho Terminal}
	\label{fig:subfigureQtdeCont_b}
\end{subfigure}
\caption{Vehicular Connectivity for two regions of Curitiba (afternoon).}
\label{fig:subfigureQtdeCont}
\end{figure}

\noindent
\emph{Vehicular Connectivity:} 
Figure \ref{fig:subfigureQtdeCont} shows vehicle connectivity (amount of buses within transmission contact) considering the two regions in the afternoon, and the two transmission range. It can be noticed that near Rui Barbosa Square (top plot),
for the 150 m transmission range, half of the records resulted in only one bus per contact, with a small variability between 1 and 2 buses. The same happened in the Pinheirinho Terminal region (bottom plot).
Both graphs presented very similar distributions, with a higher vehicular connectivity in the central region (up to 14 buses) than in the southern region (up to 6 buses). Considering the transmission range of  300 m, both regions presented practically the same data variability (between 1 and 3 buses per contact and median of 2 buses). 
	That shows that, concerning connectivity, both regions have the same characteristics, with low number of buses per contact, i.e., buses are rather far away from each other. \\

\begin{figure}[h]
\centering
\begin{subfigure}[t]{.35\textwidth}
	\centering
	\includegraphics[width=\textwidth]{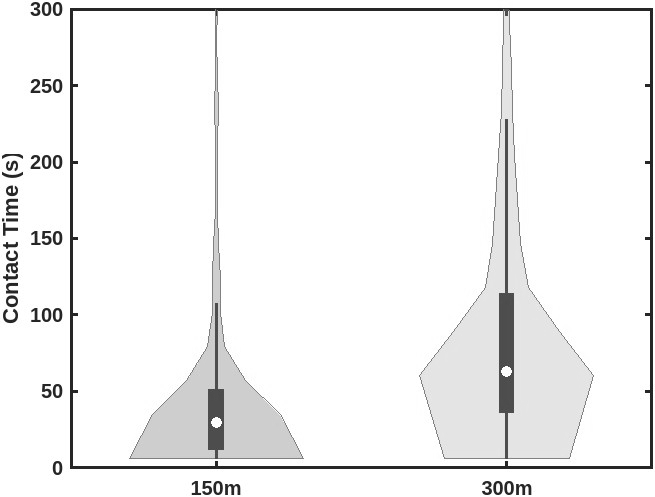}
	\caption{Contact Time}
	\label{fig:subfigureTempContRuiBarbosa_a}
\end{subfigure}
\qquad
\begin{subfigure}[t]{.35\textwidth}
	\centering
	\includegraphics[width=\textwidth]{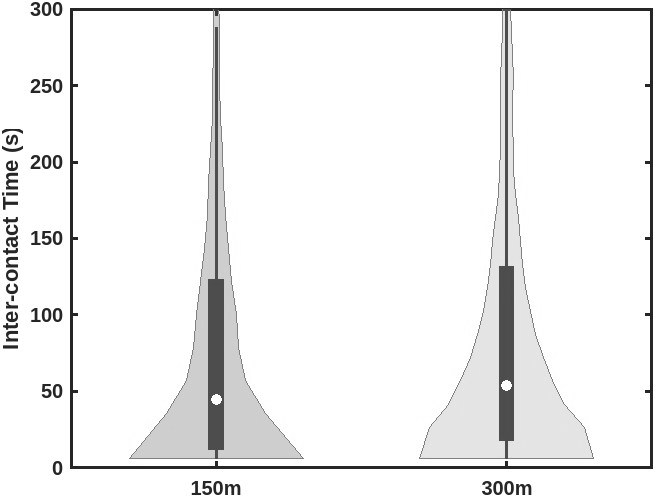} 
	\caption{Inter-contact Time}
	\label{fig:subfigureTempContRuiBarbosa_b}
\end{subfigure}
\caption{Contact Time and Inter-contact Time for Rui Barbosa Square (afternoon).}
\label{fig:subfigureTempContRuiBarbosa}
\end{figure}

\begin{figure}[h]
\centering
\begin{subfigure}[t]{.35\textwidth}
	\centering
	\includegraphics[width=\textwidth]{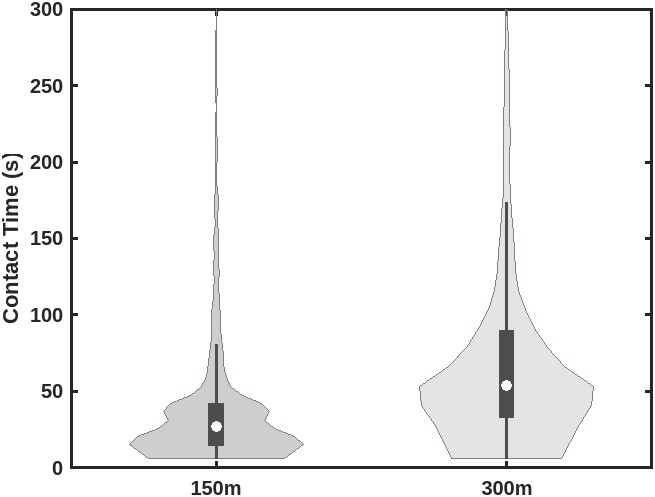}
	\caption{Contact Time}
	\label{fig:subfigureTempContPinheirinho_a}
\end{subfigure}
\qquad
\begin{subfigure}[t]{.35\textwidth}
	\centering 
	\includegraphics[width=\textwidth]{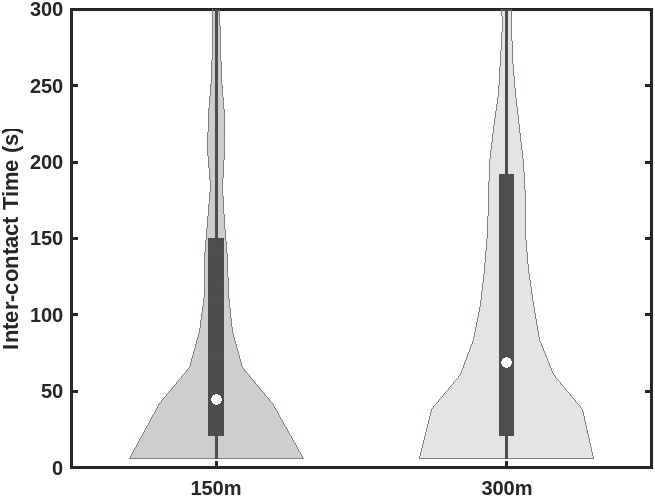} 
	\caption{Inter-contact Time}
	\label{fig:subfigureTempContPinheirinho_b}
\end{subfigure}
\caption{Contact Time and Inter-contact Time for Pinheirinho Terminal (afternoon).}
\label{fig:subfigureTempContPinheirinho}
\end{figure}

\noindent
\emph{Contact Time and Inter-contact Time:}
Figures \ref{fig:subfigureTempContRuiBarbosa} and \ref{fig:subfigureTempContPinheirinho} show results that enable to assess the contact dynamics between buses in the central and southern regions of the city. The charts show the vehicle contact time (uninterrupted intervals when buses have at least one neighboring vehicle within their transmission range) and the inter-contact time (uninterrupted intervals when the buses are isolated), considering the two examined transmission ranges. 
For the sake of a better visualization, the graphics have been limited to a maximum of 300 seconds.
It can be noticed that the distribution of these metrics exhibits, in general, a long tail.
Figure \ref{fig:subfigureTempContRuiBarbosa}  (central region) shows short connection times if compared to inter-contact time. For the 150 m transmission range, the mean contact time is approximately 57 seconds and the inter-contact time is 132 seconds (more than twice the contact time). With an increase in the transmission range, also the connection times increases, as expected. The inter-contact time, on the other hand, undergo no significant changes.
The same contact dynamics between buses is present in the southern part, as can be seen in Figure \ref{fig:subfigureTempContPinheirinho}, in which the distributions presented are very close to those of the central area, although with a smaller dispersion and lower values in contact time when compared to inter-contact time. For the 150 m transmission range, the mean contact time was about 40 seconds and the mean inter-contact time approximately 145 seconds. For the 300 m transmission range, the mean contact time was close to 107 seconds and 163 seconds for the mean inter-contact time, in addition to a greater data variability. 

\section{Conclusion} 
\label{sec:6} 

This work proposed a methodology for modeling and analyzing bus-based networks. Understanding mobility in mobile networks includes examining how entities move along and their roles in network connectivity. A large number of design issues such as routing, content dissemination or resource management much depend upon what one expects in terms of node mobility. Through the use of EMMS, this work conducted an exploratory study involving bus lines belonging to the urban transportation system of the city of Curitiba, considering central and southern regions (Rui Barbosa Square and Pinheirinho Terminal). 

Comparison between simulated and real data showed that EMMS was efficient in the process of map-matching. Simulations in SUMO showed a compatibility degree, respect to the travel times of the vehicles in the real and simulated environment, of 91.17\% for the central region and of 90.42\% for southern region of the city.

The analysis showed that the contact times between the buses are generally of short duration, on the order of 50 seconds. The intervals between contacts, on the other hand, are about twice to three times higher than contact times (from 100 to 150 seconds). This indicates the formation of small and temporary groups of vehicles (with averages between 2 and 3 buses per group) which can allow data exchange between themselves.
It is noteworthy that results obtained for other dates and time periods (omitted here due to space limitations) show that the shape of the distributions (although with possibly different means and variances), associated with each metric, maintains very similar patterns.

Given the dynamic character of the network studied in this paper, it is possible to state that the possible types of communication services allowed in such a network would be classified as opportunistic and delay-tolerant. 
Accordingly, as subject of future work, our model can be further improved by incorporating a network simulator (such as NS-2 or Omnet++)
in order to analyze the performance of the main routing protocols for opportunistic and delay-tolerant networks.




%
%
\bibliographystyle{IEEEtran}
\bibliography{reflatex} 

\end{document}